\documentclass[twocolumn,showpacs,preprintnumbers,amsmath,amssymb]{revtex4}
\usepackage{graphicx}% Include figure files
\usepackage{dcolumn}% Align table columns on decimal point
\usepackage{bm}% bold math
\usepackage[dvips]{epsfig}

\begin{document}

\preprint{Physical Review E, in Press}

\title{Diffusion of a granular pulse in a rotating drum}

\author{Nicolas Taberlet$^{1}$ and Patrick Richard$^2$}
\affiliation{1. DAMTP, University of Cambridge, Wilberforce Road, Cambridge, CB30WA, U.K.\\2. GMCM, Universit\'e Rennes 1, CNRS UMR 6626, Bat 11A, 35042 Rennes, France}

\date{\today}

\begin{abstract}
The diffusion of a pulse of small grains in an horizontal rotating
drum is studied through discrete elements methods simulations. We present a theoretical
analysis of the diffusion process in a one-dimensional
confined space in order to
elucidate the effect of the confining end-plate of the drum. We then
show that the diffusion is neither subdiffusive nor superdiffusive but
normal. This is demonstrated by rescaling the concentration profiles
obtained at various stages and by studying the time evolution of the
mean squared deviation. Finally we study the
self-diffusion of both large and small grains and we show that it is
normal and that the diffusion coefficient is independent of the grain size. 
\end{abstract}

\maketitle

\section{Introduction}

One of the most surprising features of mixtures of grains of different size,
shape or material is their tendency to segregate (i.e., to unmix) under
a wide variety of conditions~\cite{Makse1998,Knight1993,Duran2000,Ristow2000}.
Axial segregation has been extensively studied experimentally~\cite{Oyama1939,Hill1995,Hill1997a,Choo1997,Choo1998,Ristow1999,Shinbrot2000,Fiedor2003,Newey2004,Khan2004,Newey2005,Taberlet2005b,Khan2005},
numerically~\cite{Shoichi1998,Rapaport2002,Taberlet2004,Taberlet2005a} and theoretically~\cite{Savage1993,Zik1994,Levitan1997,Levine1999,Aranson1999,Elperin1999}. Yet, full understanding is still lacking.
Axial segregation occurs in an horizontal rotating drum partially
filled with an inhomogeneous mixture of grains. This phenomenon (also
known as banding) is known to perturb
industrial processes such as pebble grinding or powder mixing. After typically a
hundred rotations of the drum the grains of a kind gather in
well-defined regions along the axis of the drum, forming a regular
pattern. When the medium consists of a binary mixture of two species
of grains differing by their size, bands of small and large particles
alternate along the axis of the drum. The bands of small grains can be
connected through a radial core which runs throughout the whole
drum~\cite{Cantelaube1995,Hill1997b}. The formation of the radial core occurs quickly in the
first few rotations as the small grains migrate under the surface.\\

Unlike radial segregation, axial segregation requires a transport of
grains along the axis of the drum. Since the grains are initially
mixed, they must travel along the axis of the drum to form bands. This
underlines the importance of the transport mechanisms in a rotating
drum.
Much theoretical work as been devoted to axial segregation, most of
which assumes a normal diffusion of the grains along the axis~\cite{Savage1993,Zik1994,Levitan1997,Levine1999,Aranson1999,Elperin1999}.
This hypothesis has recently been challenged by Khan {\it et
  al.}~\cite{Khan2005} who have reported remarkable experimental
  results. These authors studied the diffusion of an initial pulse of
  small grains among larger grains in a long drum. 
Since direct visualisation is not possible (because the small grains
are buried under the surface), these authors used a projection
  technique. Using translucent large grains and opaque small grains,
  they recorded the shadow obtained when a light source is placed behind
  the drum.
They found the diffusion process to be subdiffusive and to scale
  approximatively as $t^{1/3}$.
They also studied the self-diffusion of salt grains and found again a
sub-diffusive process with an similar $t^{1/3}$ power law.
In this article, we report numerical findings on the diffusion of a
pulse of small grains. Interestingly, our results are in
contradiction with those of Khan {\it et al.}~\cite{Khan2005} since we
observed normal diffusion.

\begin{figure}[h]
\begin{center}
\includegraphics*[width=6cm]{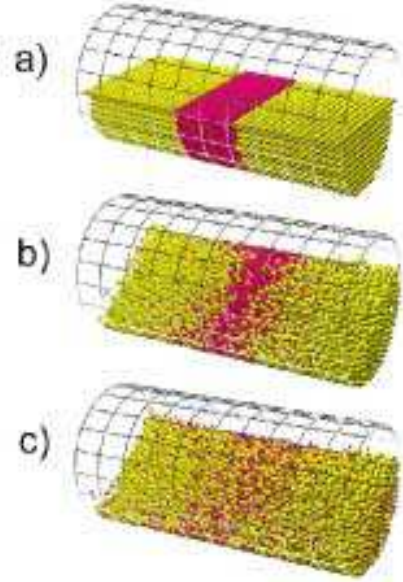}
\caption{(Color online) Snapshots of the DEM simulation taken
  after 0 rotations (a), 30 rotations (b) and 100 rotations (c).
A initial pulse of small (red oneline) grains is inserted a the center of a
  rotating drum otherwise  partially filled with large grains (yellow oneline).
For clarity we used here an initial pulse of length $l=  25d_S$ in a
drum of length $L=120d_S$. Note that longer drums are used later.}
\label{fig_diff2}
\end{center}
\end{figure}

The outline of the paper is as follow. First we will describe the
simulation method. A theoretical analysis of one-dimensional diffusion in a
confined space is then presented. We present concentration profiles
and mean squared deviation and show that the diffusion process is neither
subdiffusive or superdiffusive. An instability in the average position is
described. Finally, we report results on the self-diffusion of both small
and large grains.

\section{Description of the simulation}

This article presents results based on the soft-sphere molecular dynamics (MD) method,
one of the Discrete Elements Methods (DEM). This method deals with
deformable frictional spheres colliding with one
another. Although not flawless, it has been widely used in
the past two decades and has proven to be very reliable~\cite{Schafer1996,Taberlet2004}.
Here we study the diffusion of a pulse of small grains among larger
grains in an horizontal rotating drum (see Fig. 1).
We follow the positions of individual small grains and in the
following, it is
understood that we constantly refer to the small grains. In
particular, the ``concentration'' means concentration of small grains.

\subsection{Parameters of our simulations}

The mixture consists of two species of ideally spherical grains differing by their
size. The small grains have a diameter $d_S = 5$ mm and the large
grains have a diameter of $2d_S$. The density is the same for both kinds of beads: 
$\rho = 0.6 \, \mbox{g}/\mbox{cm}^3$.
The length of the drum, $L$, is varied from $L = 60 d_S$ to $L = 420d_S$ and its radius is set to $20\,d_S$. The rotation speed is set to
$0.5 \, \mbox{rot}\mbox{s}^{-1}$. 
The grains are initially placed in a cubic grid. The small grains are placed in the middle of the drum around
$x=0$ and the large grains fill up the space between the pulse of
small grains and the end plates. The number of small grains can be varied but unless otherwise
mentioned is set to $N=10\,300$, which corresponds to an initial pulse
of length $l=25d_S$. The number of large grains depends of course on
the length of the drum. The simulations run for
typically a few hundreds of rotations.

The rotation is started and the medium compacts to lead to an average
filling fraction of 37\%. Radial segregation occurs rapidly
(typically after five rotations of the drum) and the initial pulse is
buried under the surface.
The concentration profiles can be obtained at any time step. Knowing the exact
position of every grain allows one to accurately
compute the average position $\langle x \rangle$ and the mean squared deviation $\langle x^2 \rangle$.
Note that periodic boundary conditions could be used in order to
avoid the effect of the end plates. However, the
confinement would still play a role since the position of individual
grains would still be limited. Moreover it may lead to nonphysical spurious effects.

\subsection{The molecular dynamics method}

The forces schemes used are the dashpot-spring model for the normal
force $F_{ij}^n$ and the regularized Coulomb solid friction law for
the tangential force~\cite{Frenkel96} $F_{ij}^t$: respectively, $
F_{ij}^n = k_{ij}^n \delta_{ij} -\gamma_{ij}^n  \dot{\delta}_{ij}$ and
$F_{ij}^t = \mbox{min}(\mu F_{ij}^n, \gamma^t v^s_{ij})$,  where $\delta_{ij}$
is the virtual overlap between the two particles in contact defined
by: $\delta_{ij} = R_i + R_j - r_{ij}$, where $R_i$ and $R_j$ are the
radii of the particles $i$ and $j$ and $r_{ij}$ is the distance
between them . The force acts whenever
$\delta_{ij}$ is positive and its frictional component is oriented in
the opposite direction of the sliding velocity. $k_{ij}^n$ is a spring
constant, $\gamma_{ij}^n$ a viscosity coefficient producing
inelasticity, $\mu$ a friction coefficient, $\gamma^t$ a regularization
viscous parameter, and $v^s_{ij}$ is the sliding velocity of the contact.
If $k_{ij}^n$ and $\gamma_{ij}^n$ are constant, the restitution coefficient,
$e$, depends on the species of the grains colliding.
In order to keep $e$ constant the values of $k_{ij}^n$ and $\gamma_{ij}^n$
are normalized using the effective radius
$R_{\mbox{eff}}$
defined by 
$1/R_{\mbox{eff}} = 1/R_i + 1/R_j$: $k_{ij}^n = k_0^n R_0 /R_{\mbox{eff}}$
and $\gamma_{ij}^n = \gamma_0^n R_{\mbox{eff}}^2/R_0^2$.
The particle/wall collisions are treated in the same fashion
as particle/particle collisions, but with one particle having infinite mass and radius.
The following values are used: $R_0=4$ mm, 
$k_0^n=400\;\mbox{Nm}^{-1}$, %$\gamma_0^n=0.012\;\mbox{kg}\mbox{s}^{-1}$
(leading to $e \simeq 0.9$), $\gamma^t=6\;\mbox{kg}\mbox{s}^{-1}$ and
$\mu = 0.3$.
The value of $e$ was varied
(from 0.4 to 0.9), which seemed to have only very little influence on the diffusion process.

The equations of motion are integrated using the Verlet method with a
time step $dt=1/30\,\Delta t$, where $\Delta t$ is the duration of a
collision ($\Delta t \approx 10^{-3}$s). The simulations are typically
run for $10^7$ time steps, corresponding to a few hundreds of rotations.

\section{1D confined diffusion}
Before analyzing any results, one should quantify the influence of the
end-plates. The limited space imposes some constraints on the diffusion
process: the position $x$ along the axis of the drum can only range
from $-L/2$ to $L/2$ which imposes a limit to the mean squared deviation.
In order to clarify the effect of the confinement we present in this
section a theoretical analysis of an initial Dirac distribution
(corresponding to the initial pulse) diffusing in a confined
1D space. Note that an initial pulse function could also be used. However, this
would add a degree of complexity to the problem without much
benefit. In this brief section we only intend to qualitatively
describe the effect of the confinement rather than studying it in details. The concentration, defined for $x \in [-L/2,L/2]$, is given by~\cite{Carslaw59}:

$$
c(x,t) = \displaystyle \frac{1}{\sqrt{4\,\pi\,D\,t}}\; \displaystyle
\sum_{n=-\infty}^{\infty}
\exp^{\displaystyle \frac{-(x+n\,L)^2}{4\,D\,t}}.
$$
Note that the exact solution for an initial pulse of nonzero width can be
obtained by convoluating this solution to the initial pulse function.

\begin{figure}[htbp]
\begin{center}
\includegraphics*[height=3.2cm]{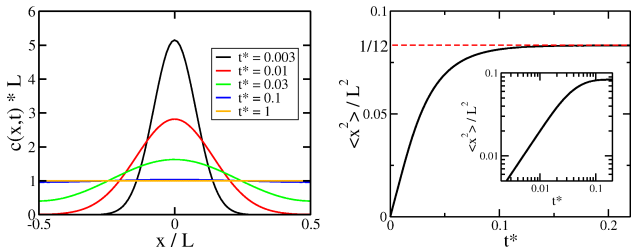}
\caption{(Color online) Left: concentration profiles obtained at various
  stages for a confined 1D diffusion process. Right: mean squared deviation versus dimensionless time. At short times the
  function is a linear function of time but at longer time $\langle
  x^2 \rangle$ saturates. Inset: the data plotted on a log-log
  scale.}
\label{fig_Pxt}
\end{center}
\end{figure}

We can now plot the concentration profiles for various times (see
Fig.~\ref{fig_Pxt}). Note that the time and position can be made dimensionless
(using $t^* = t \, L / D^2$ and $x^*=x/L$).
One can see that at short time the distribution is almost a Gaussian
but at longer times, the confinement plays a important role. The distribution
flattens and eventually reaches a constant value of $c(x) = 1/L$,
which corresponds to a mean squared deviation $\langle x^2 \rangle_{\infty} = L^2/12$.
Figure~\ref{fig_Pxt} also shows the time evolution of the mean
squared deviation defined here by $\langle x^2 \rangle = \int_0^L
c(x,t) \, x^2 \, dx$. The
function is clearly linear at short time (as confirmed by the log-log
scale inset) but saturates a longer times. This behavior was expected
but can lead to erroneous conclusions. Indeed, the curvature observed
in $\langle x^2 \rangle (t)$ could be mistaken for a sign of a
subdiffusive process whereas it is solely due to the confinement.
Note that Khan {\it et al.}~\cite{Khan2005} used a rather long
tube in their experiments, meaning that the confinement did not affect
their results.

\section{Results}
\subsection{Concentration profiles}

\begin{figure}[t]
\begin{center}
\includegraphics*[width=7.3cm]{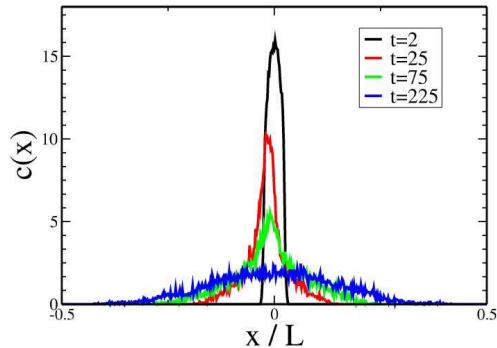}
\caption{(Color online) Concentration profiles at different time obtained with a
  drum of length $L= 420 \,d_S$.}
\label{fig_cx_z4}
\end{center}
\end{figure}

We now present results obtained in our 3D numerical simulations.
The position of every grain at any time step is known, which allows one
to compute the concentration profiles. The drum is divided in virtual
vertical slices of length $d_S$ and the number of grains whose center
is in a given slice is computed. Figure~\ref{fig_cx_z4} shows concentration profiles of
small grains measured at short times for a drum of length $L = 420 \,
d_S$. Note that the profiles tend toward
Gaussian distributions as the grains slowly diffuse in the drum. 

\begin{figure}[b]
\begin{center}
\includegraphics*[width=8.4cm]{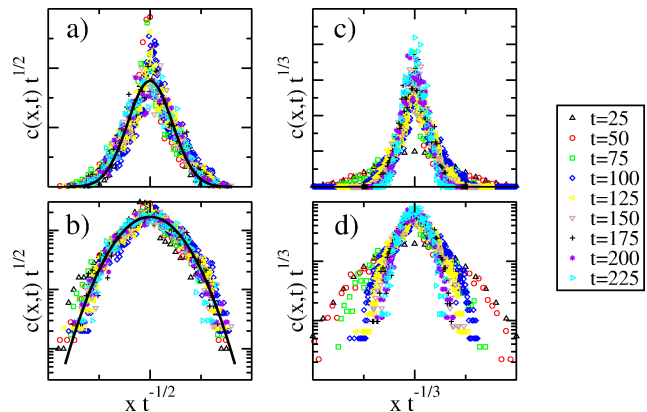}
\caption{(Color online) Rescaled concentration profiles (in arbitrary units) at different times obtained with a
  drum of length $L= 420 \,d_S$. The data is rescaled using $t^{1/2}$
  for (a) and (b), and $t^{1/3}$ for (c) and (d). (a) and (c): linear
  scale, (b) and (d):  semilog scale. The solid line on (a) and (b) is a
  Gaussian fit at $t=225$ rotations. On can see that the $t^{1/2}$
  rescaling leads to a good collapse unlike the $t^{1/3}$
  rescaling. This is particularly clear on the semilog plots.
}
\label{fig_Nb_collaps}
\end{center}
\end{figure}

Note that no axial segregation is visible at any point. Indeed, except
for the obvious one at $x=0$, there is no peak in the
concentration. We believe that the number of small grains is too small
compared to that of large grains to trigger axial segregation.\\

A very good test to check whether the diffusion process is
subdiffusive, normal or superdiffusive is to try to collapse various
concentration profiles at different times onto one unique curve. This
should be done by dividing the position by some power of the time
($t^\alpha$) and for reasons of normalization (i.e. mass conservation)
by simultaneously multiplying the concentration by $t^\alpha$. The
process is subdiffusive if $\alpha < 1/2$, normal if $\alpha = 1/2$
and super-diffusive if $\alpha > 1/2$. Using the profiles taken at
``short times'' (i.e., before the confinement plays a role), we were able to obtain an excellent collapse of our data
using the value $\alpha = 1/2$. Figures~\ref{fig_Nb_collaps}(a) and \ref{fig_Nb_collaps}(b) shows the data in a
linear and semilog scales. Figures~\ref{fig_Nb_collaps}(c) and
\ref{fig_Nb_collaps}(d) shows similar plots obtained using $\alpha=1/3$,
as suggested by Khan {\it et al.}~\cite{Khan2005}.
The first conclusion that can be drawn is
that the collapse is excellent for $\alpha=1/2$ and poor for
$\alpha=1/3$. This is particularly clear on the semilog plots. The
profiles on Fig.~\ref{fig_Nb_collaps}(d) tend to narrow with increasing
time whereas no trend is visible on Fig.~\ref{fig_Nb_collaps}(b).
This clearly shows that in our simulations the diffusion is a normal
diffusion process.
Moreover, the collapsed data can be
well fitted by a Gaussian distribution, as clearly evidenced by the
semilog plot on Fig.~\ref{fig_Nb_collaps}(b).

\subsection{Mean squared deviation}

\begin{figure}[h]
\begin{center}
\includegraphics*[width=6.4cm]{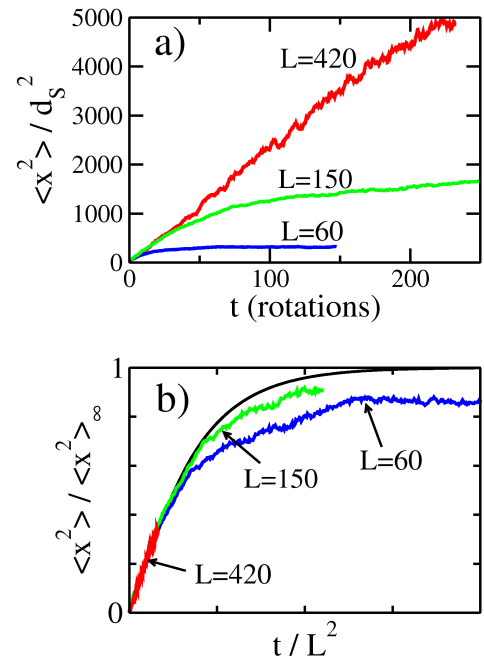}
\caption{(Color online) (a) Mean squared deviation versus time (in arbitrary units). (b) Rescaled data. The
  different curves correspond to different values of the drum
  length: Blue oneline: $L=60 \, d_S$, green online $L=150 \, d_S$ red oneline : $L=420 \,
  d_S$. The solid line in (b) is the theoretical prediction.}
\label{fig_x2_simu}
\end{center}
\end{figure}

The collapse of the different curves on Fig.~\ref{fig_Nb_collaps} is
a good indication that the diffusion process in normal. An even better
test is to plot the time evolution of the mean squared deviation
defined here by $\langle x^2 \rangle = 1/N \, \sum_i (x_i-x_i^0)^2$
where $x_i^0$ is the initial position of the grain number
$i$. Figure~\ref{fig_x2_simu}a is a plot of the mean squared deviation
versus time obtained for various values of $L$. Of course the
saturation value $\langle x^2 \rangle_{\infty}$ is different
for different values of $L$ but one can see that the initial slope is
the same for all curves. This shows that the diffusion coefficient is
well defined at ``short times'' and does not depend on the length of the drum. Figure~\ref{fig_x2_simu}(b) is a rescaled
plot of the same data. The time is divided by $L^2$ and the mean
squared deviation by $\langle x^2 \rangle_{\infty}$. One can see that
all data collapse at short times, showing again that the
diffusion coefficient is well defined and that it is independent of
$L$.

\begin{figure}[htbp]
\begin{center}
\includegraphics*[width=6.4cm]{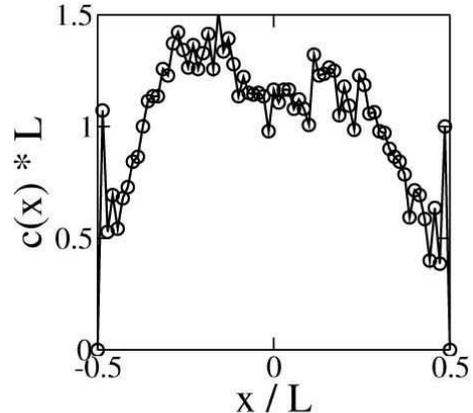}
\caption{(Color online) Steady state concentration profile for $L= 60d_S$ after 150
  rotations. The distribution is not perfectly uniform: the
  concentration in small grains drops near the end plates.}
\label{fig_steady_state}
\end{center}
\end{figure}

One can see on Fig.~\ref{fig_x2_simu}(b) that the mean squared deviation
does not reach the theoretical saturation value $\langle x^2 \rangle_{\infty}$. This is
an indication that in the final state the concentration is not
uniform. Indeed although no segregation bands are visible there is
still a signature of axial segregation. The concentration profile in
the steady state obtained for $L=60 \, d_S$ is shown on
Fig.~\ref{fig_steady_state}. One can see that the initial central peak
has diffused but the concentration is not even in the final state: the
concentration drops near the end plate, which has also been observed
experimentally. This explains why the saturation value $\langle x^2 \rangle_{\infty} = L^2/12$ is not reached
in our simulations. Note, however, that for longer drums this effect becomes negligible.\\

\begin{figure}[h]
\begin{center}
\includegraphics*[width=8.4cm]{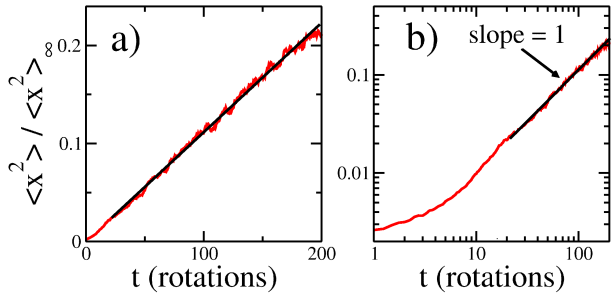}
\caption{(Color online) Mean squared deviation versus time in a linear (right) and
  log-log (left) scale for $L=420d_S$. Except for a
  brief transient during which radial segregation occurs $\langle x^2
  \rangle$ is a linear function of time the data plotted in a log-log
  plot scale is extremely
  well fitted by a line of slope 1.}
\label{fig_x2_z4}
\end{center}
\end{figure}

The rescaled plot on Fig.~\ref{fig_x2_simu}b allows one to decide which
time frame should be considered ``short times''. In particular it
is obvious from that picture that the run performed with $L=420 \,
d_S$ has not yet been affected by the end-plates after 200
rotations. Therefore, it allows one to study the diffusion process
without considering the consequences of the
confinement. Figure~\ref{fig_x2_z4} shows the mean squared
deviation versus time in linear and log-log scales for a drum
length $L=420 \,d_S$. Both plots clearly demonstrate that the
diffusion is normal. Indeed, the data on Fig.~\ref{fig_x2_z4} is
well-fitted by a straight line on a linear scale. Similarly, after a short transient
corresponding to the time it takes for radial segregation to be
completed (approximatively 10 rotations here), the data plotted in a
log-log scale is perfectly fitted by a line of slope 1 for over a
decade. These
observations show without a doubt that the diffusion process in our
simulations is neither subdiffusive nor superdiffusive but simply normal.

\section{Self-diffusion}

In this section we will study the self-diffusion process. We would like to
compare results obtained with three runs. The first run consists of
monodisperse small particles, the second of large particles and the
third one is the pulse experiment described above. The rotation speed,
filling ratio and drum size are identical all three cases. In order
to save computational time, we used a rather short drum ($L=60 d_S$).
The major difference between a self-diffusion numerical experiment and the
diffusion of a pulse is that no radial segregation can exist in the
self-diffusion. Since the mixture is monodisperse there cannot exist
any form of segregation. It is therefore interesting to compare the two
numerical experiments and elucidate the role of the radial segregation
in the diffusion process.\\

What should be measured in a self-diffusion experiment? Of course, there
is no initial pulse in the drum but one can be arbitrarily
defined. Indeed, one can track the grains whose initial position were
within a given distance from the center of the drum. This would allow
to plot ``virtual concentration profiles.'' More simply, we can use
the definition used before for the mean squared deviation $\langle
x^2 \rangle = 1/N \, \sum_i (x_i - x_i^0)^2$. 
Note, however, that the end plates play the same role as before. Indeed,
the position is still limited. Moreover, the grains initially located
near an end plate will ``feel'' the end plates at early stages. It is
therefore necessary to measure the mean squared deviation of grains
initially centered around $x=0$. For both self-diffusion
simulations we used the grains located within a distance of $10 d_S$
from the center of the drum ($x=0$).

\begin{figure}[h]
\begin{center}
\includegraphics*[width=6.4cm]{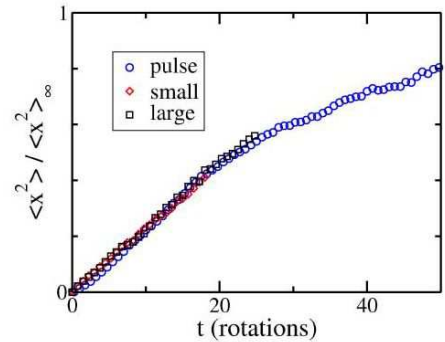}
\caption{(Color online) Mean squared deviation versus time for the
  pulse experiment (blue oneline),
  self-diffusion of small (red oneline) and large (black) grains. One can see
  all three curves collapse, showing that the self-diffusion is normal
  and that the diffusion coefficient is independent of the grain size.}
\label{fig_self}
\end{center}
\end{figure}

Figure~\ref{fig_self} is a plot of  $\langle x^2 \rangle$ versus time
for all three runs. Rather surprisingly, all three curve
collapse. This shows that the diffusion process is independent of the
grain size, which is very interesting. One could expect the diffusion
coefficient to scale with the grain diameter and the two self-diffusion
coefficients to be different but it is obviously not the
case. Instead, our simulations show that $d_S$ is not a relevant
length scale regarding the diffusion process.

Maybe even more interestingly, the data collapse shows that radial
segregation does not seem to affect the diffusion process. Indeed, the
diffusivity is the same in the pulse experiment (where radial
segregation exist) as in the self-diffusion experiments (where no
segregation can exist).

\section{Discussion and conclusion}

The diffusion of an initial pulse seems to be very helpful in
understanding the basic mechanisms leading to axial segregation. 
It is however difficult to obtain precise measurements in an
experimental system. Khan {\it et al.}~\cite{Khan2005} developed a clever
projection technique which allows one to observe the hidden core of
small grains. The shadow projected by the opaque small grains is
clearly related to the concentration in small grains but one can
question whether the link between the two is a linear relation or a
more complex one.

\begin{figure}[t]
\begin{center}
\includegraphics*[width=8.4cm]{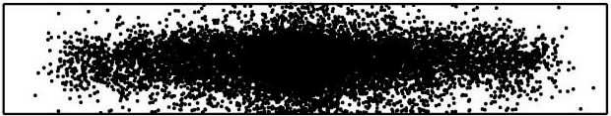}
\caption{View of the small grains projected perpendicularly to the
  free surface for $L=280d_s$ after 75 rotations (i.e., before the
  confinement plays a role).}
\label{fig_core}
\end{center}
\end{figure}

Our numerical simulations show consistent results indicating that the
diffusion is normal. This conclusion is strongly supported by a collapse of
concentrations profiles rescaled by $t^{1/2}$. Moreover, the mean squared
deviation is clearly a linear function of time as long as the
confinement does not play a role. These results are in strong
contradiction with those obtained experimentally by Khan {\it et
  al.}~\cite{Khan2005}. These authors found a subdiffusive process which
  scales approximatively as $t^{1/3}$. The origin of this discrepancy
  remains unknown.

It could originate from the force models used in
our simulations. In particular, it would be very interesting to test
the consistency of our results using the Hertz modelor a tangential spring
friction model~\cite{Schafer1996} .
The discrepancy might also originate in the projection method used by
Khan {\it et al.}~\cite{Khan2005} which is an indirect measurement of the concentration. However, it seems unlikely that the details of the projection techniuqe would change the
diffusion power laws.
One interesting check for that method would be
to apply the projection technique to our numerical data. 
Figure~\ref{fig_core} shows a projection
(perpendicularly to the free surface) of the core of small
grains obtained for $L=280d_s$ after 75 rotations. This image looks
rather differents from those of Khan {\it et al.}~\cite{Khan2005}. In particular, one
can observe individual grains, which is not possible using their
projection technique. 
Applying the projection technique to our numerical data
is not straightforward and the definition of such a procedure would obviously drastically influence the outcome.
Note, however, that Khan {\it et al.} have also observed a 
subdiffusive self-diffusion using direct imaging, which does not involve projection. This supports the idea that the 1/3 power law which they reported is not an artifact of their projection technique and that the origin of the discrepancy between the experiments and the simulations may be due to some unaccounted for physical effect.

After stimulating discussions with these authors, we believe that
there are a number a physical differences between our two system that
might lead to different results.
One difference between the two systems is the ratio of the drum
diameter to the avarage particle diameter $\delta = D/\bar{d}$. In
our simulations $\delta=26$ and in their experiments $\delta \approx 100$. 
The discrepancy might also be caused by the difference in the shape of the
grains although a subdifussive process was observed experimentally using bronze beads.
Finally, the experiments were conducted in a humidity-controlled room and the capillary bridges between the grains might have modified the diffusion process.\\

In conclusion, our numerical results showed that the diffusion along
the axis of the drum is normal. Having studied the effect of the
confinement, we can conclude that the curvature observed in the mean
squared deviation is not a sign of subdiffusion. The rescaled
concentration profiles lead to the same conclusion that the diffusion
process is normal. The study of self-diffusion shows that the diffusivity is independent of the grain size and is not
affected by radial segregation. This results may shed some light on the mechanisms of
formation of segregation bands since it indicates that the transport of grains along the
axis of the drum is identical for both species of grains. We hope that
these results will inspire theoretical models and help understanding
the puzzling phenomenon of axial segregation.

\section{Acknowledgements}
The authors would like to thank M. Newey, W. Losert, Z. Khan and S.W. Morris for 
fruitful discussions.

%\bibliographystyle{apsrev}    % <---------------------------------- MOD
%\bibliography{diffusion}

\end{document}